\documentclass[aps,twocolumn,prc]{revtex4}
\usepackage[dvips]{graphicx}
\usepackage{amsmath}
\usepackage{amssymb}
\begin{document}
\title{Mass, radius, and composition of the outer crust of nonaccreting cold neutron stars}
\author{Matthias Hempel}
\email{hempel@astro.uni-frankfurt.de}
\affiliation{Institut f{\"u}r Theoretische Physik, J. W. Goethe-Universit{\"a}t, 60438 Frankfurt am Main, Germany}
\author{J{\"u}rgen Schaffner-Bielich}
\email{schaffne@astro.uni-frankfurt.de}
\affiliation{Institut f{\"u}r Theoretische Physik, J. W. Goethe-Universit{\"a}t, 60438 Frankfurt am Main, Germany}
\begin{abstract}
The properties and composition of the outer crust of nonaccreting cold neutron stars are studied by applying the model of Baym, Pethick, and Sutherland, which was extended by including higher order corrections of the atomic binding, screening, exchange and zero-point energy. The most recent experimental nuclear data from the atomic mass table of Audi, Wapstra, and Thibault from 2003 is used. Extrapolation to the drip line is utilized by various state-of-the-art theoretical nuclear models (finite range droplet, relativistic nuclear field and non-relativistic Skyrme Hartree-Fock parameterizations). The different nuclear models are compared with respect to the mass and radius of the outer crust for different neutron star configurations and the nuclear compositions of the outer crust.
\end{abstract}
\maketitle
\section{Introduction}
In 1971 Baym, Pethick, and Sutherland (BPS) determined in their classical paper \cite{BPS} the equation of state and the sequence of nuclei appearing in the outer crust of cold nonaccreting neutron stars. They used a droplet model of Myers and Swiatecki \cite{MyersSwiatecki} for the masses of the nuclei, the only input informationen needed in this calculation. With time, more and more masses of so far unknown nuclei became available. Also theoretical models with increased precision were developed. Several publications followed \cite{HZD,HP,RHS,Guo}, which updated the results of BPS by applying the newest experimental data and theoretical mass tables available at that time. In the present publication we take a selection of the theoretical nuclear mass tables and results presented in \cite{RHS} and analyze the impact of the different models on the mass radius relation and the compositions of the outer crust for different realistic neutron star configurations. 

Our study of the outer crust is based on the BPS-model, extended by including higher order corrections. The only input needed are the masses of the nuclei. Therefore, we take the most recent experimental data available, the 2003 atomic mass table from Audi, Wapstra, and Thibault \cite{AW}. To obtain nuclear masses, the atomic electron binding energies were subtracted with an empiric formula as given in \cite{lunney}. For nuclei with unknown masses, five different theoretical nuclear models were used. BSk8 is a Skyrme-Hartree-Fock-Bogolyubov (Skyrme-HFB) calculation \cite{bsk8}. SkM* is a Skyrme-Hartree-Fock model with a BCS pairing force \cite{skm}. FRDM stands for the most recent version of the finite-range droplet model \cite{frdm}. NL3 is a well known parameterization of the relativistic mean field model \cite{nl3}. The chiral Lagrangian used for the set Chiral is built on the nonlinear realization of the chiral SU(3)$\times$SU(3) symmetry as motivated from quantum chromodynamics \cite{chiral}. Besides the Chiral model, the nuclei were calculated allowing for axial deformations. The nuclear models used are described in more detail in \cite{RHS}.

The outer crust of nonacreeting cold neutron stars is built of the groundstate of cold ($T=0$), neutral and beta-equilibrated matter at densities between $10^4$ and $10^{11}$ g/cm$^3$. In this density region the distance between atoms is small compared to their spatial extension which leads to complete ionization of the atoms. The negatively charged electrons separate from the positive nuclei and can be described as a free Fermi-gas in first approximation. The nuclei arrange in a way to minimize the Coulomb interaction between themselves and the surrounding electron gas and thus form a solid (body centered cubic) lattice. 

The total pressure of the system $P$ has a contribution from the electrons $P_e$ and a negative pressure coming from the lattice: $P = P_e + \frac13 W_L n_N$, where $W_L$ is the lattice energy and $n_N$ the number density of nuclei. The total mass of the nuclei $W_N$ enters in the baryon chemical potential $\mu_b =  (W_N + \frac43 W_L + Z \mu_e)/A$, with the mass number $A$ and the charge number $Z$. The electron chemical potential $\mu_e$ is fixed by the condition of charge neutrality. Thus the baryon chemical potential at given pressure is only a function of the charge and mass number. To find the groundstate, one has to look for the nucleus $(A,Z)$ with minimal baryon chemical potential. A more detailed description of the BPS model is given in the original paper \cite{BPS}, or can be found in \cite{RHS}.

So far, the electrons were considered as free particles. In the present calculation, three higher order corrections to this approximation are included \cite{salpeter}. First the screening or Thomas-Fermi energy, which describes deviations of the electron distribution from uniformity due to the negatively charged ions. Secondly, the exchange energy, to take into account the Fermionic nature of the electron-electron interaction. Furthermore the zero-point motion of the nuclei in the lattice was incorporated, also to verify the stability of the solid crystal. The latter correction to the energy density turns out to be even smaller than the first two and is negligible with respect to the equation of state or the composition of the outer crust. 

With the lowest energy per nucleon of all elements, $^{56}$Fe represents the ground state nucleus up to densities of $10^6$ g/cm$^3$. At higher densities the electrons become relativistic and begin to change the position of equilibrium. Electron capture takes place, leading to more and more neutron rich nuclei. Exotic nuclei with large neutron excess appear, stabilized against $\beta$-decay by the filled electron Fermi sea. Finally, the baryon chemical potential exceeds the neutron mass. Now the nuclei are so neutron rich that some neutrons are not bound any more and start to drip out of the nucleus. The neutron drip is reached and the inner crust begins.

The equilibrium nucleus is determined by an interplay of the strong, short ranged nuclear interactions present in the nuclei and the electromagnetic forces between the charged particles. In this regard the outer crust is an fascinating astrophysical site, where theoretical nuclear models determine macroscopic, in principle observable, properties of neutron stars. 

The internal structure of a neutron star is described by the Tolman-Oppenheimer-Volkoff (TOV) equation \cite{tov}: 
\begin{eqnarray} 
 \frac{d P}{d r} &=& - \frac{G \epsilon(r) {m}(r)} {c^2 r^2} 
 \cdot \left[1 + \frac{4 \pi r^3 P(r)} {{m}(r) c^2} \right] \nonumber \\ 
 & & \cdot \left[ 1 + \frac{P(r)}{\epsilon(r)} \right]
 \cdot\left[1 - \frac{2 G {m}(r)} {c^2 r} \right]^{-1}\; ,
 \label{tov}
\end{eqnarray}
with the radius $r$, the mass $m$, the gravitational constant $G$, and the speed of light $c$. With the knowledge of the equation of state, i.e.~the energy density $\epsilon(P)$ as a function of the Pressure $P$, the mass and the radius corresponding to one certain central pressure $P_0$ can be calculated. Usually one starts the numerical integration of the TOV-equation in the center of the star, i.e. $r=0$, $m=0$, and $P=P_0$. But the TOV-equation also allow to begin at an arbitrary point inside the neutron star, $r=R_0$, $m=M_0$, and $P=P_0$, as far as realistic values for $R_0$, $M_0$ and $P_0$ are considered. This allows to calculate the mass and radius of the outer crust for a chosen inner neutron star configuration ($R_0$, $M_0$ and $P_0$), without any assumptions about the equation of state of the inner region below the outer crust. 

The mass and the radius are the most elementary and important measurable quantities of a neutron star. Precise mass and radius measurements can rule out equations of state which are in conflict with the experimental data, as most recently demonstrated in the case of the X-ray binary EXO 0748 - 676 \cite{EXO}. Astrophysical observations are a source of knowledge about the properties of dense hadronic matter. In this context it is interesting to study the consequences of the theoretical nuclear models on the mass and radius of the outer crust, which is the topic of this investigation.

On the other hand, the composition of the outer crust is important for heat transport and for Ohmic dissipation of the magnetic field, which will be explored in the last section. Furthermore, the composition and the equation of state are necessary to determine the eigenmodes of acoustic oscillations \cite{chugu} or toroidal shear modes \cite{watts}.
\section{Mass and radius of the outer crust}
Table \ref{t:mr} lists the mass and the radius calculated for three different equations of states, which are based on the theoretical mass model BSk8 and SkM*. Additionally one equation of state for BSk8 is considered, where effects from the lattice (and the higher order corrections to it) were neglected. The initial pressure for the beginning of the outer crust is chosen to be $P_0=6.6\times 10^{29}$ dyne/cm$^2$, a value equivalent to the neutron drip of these equations of state.
\begin{table} [ht]
\caption{\label{t:mr}The mass $\Delta M$ and thickness $\Delta R$ of the outer crust of a nonaccreting cold neutron star with radius $R_0=12$ km and mass $M_0=1.4{~\textnormal M_\odot}$ of the inner region (inner crust together with core), for three different equations of state. The initial pressure for the beginning of the outer crust is fixed to $P_0=6.6\times 10^{29}$ dyne/cm$^2$.}
\footnotesize
\rm
\begin{tabular}{@{}*{7}{l}}
\hline
EoS & $\Delta M$ [$10^{-5}~\textnormal M_\odot$]& $\Delta R$ [km]\\
\hline
\hline
BSk8&3.090&0.4509\\
SkM*&3.088&0.4408\\
BSk8 without lattice&3.093&0.4666\\
\hline
\end{tabular}
\end{table}

Compared to the total radius of a neutron star of 10 to 20 km the deviations in the radial extensions of the outer crusts are found to be quite small. In this respect the theoretical nuclear models only play a minor role in the mass radius relation. However it is surprising, that the differences of the two nuclear mass tables BSk8 and SkM* are almost as important as the inclusion of the lattice.

Next, the dependence of the mass and thickness of the outer crust on the radius $R_0$ and mass $M_0$ of the inner region will be analyzed. The equation of state which belongs to the mass table FRDM is being used. Figure \ref{b:mr} depicts the results for initial masses of $M_0=1.0 - 2.0$ M$_\odot$ and initial radii of $R_0=10 - 20$ km.

The mass as well as the thickness of the outer crust increases with $R_0$ for all initial masses. On the other hand neutron stars with large $M_0$ have smaller and lighter outer crusts. For the considered values of $M_0$ and $R_0$ the outer crust mass $\Delta M$ ranges from $7.8\times10^{-6}$ to $5.6\times10^{-4}$, the radius $\Delta R$ from 0.14 to 2.57 km. For $M_0=1.0$ M$_\odot$ the outer crust spans between 4.7 and 11.4 \% of the total radius $R+\Delta R$, for $M_0=2.0$ M$_\odot$ between 1.4 and 4.7 \%. In summary, the outer crust is the lighter and smaller the more compact the neutron star is.
\section{Composition}
In the following the composition of the outer crust for different neutron star configurations and equations of state is examined. The distribution of the fraction of nuclei on the total number of nuclei $N_{tot}$ for given mass number $A$ and charge number $Z$ is shown in Figure \ref{b:zs}. The theoretical nuclear models NL3, FRDM and the more exotic model Chiral were selected for comparison. 

The sequence of groundstate nuclei with increasing density for FRDM and NL3 follows mainly along the magic neutron numbers $N=50$ and $N=82$ \cite{RHS}, which leads to the bimodal distribution visible in Figure \ref{b:zs}. In all three models there is only a vanishingly small amount of $^{56}$Fe present, as this nucleus only appears at the lowest densities. On the contrary, elements which appear at high densities can give a significant contribution to the overall composition, like it is the case for $^{156}$Pd in the Chiral model. Although the groundstate nucleus are identical for all the three models up to $3\times10^{10}$ g/cm$^3$, this causes the visible differences in the total compositions. These results clarify that differences in the sequence are likely to be amplified in the composition.

With NL3 a mean mass number $\left<A\right>=105.7$ and mean charge number $\left<Z\right>=34.5$ is found for the $M_0=1.4~{\textnormal M_\odot}$ and $R_0=10$ km neutron star, $\left<A\right>=108.3$ and $\left<Z\right>=36.0$ for FRDM, and $\left<A\right>=104.8$ and $\left<Z\right>=34.0$ for the set Chiral. These values are relatively similar, in average the groundstate nuclei are Selenium ($Z=34$) or Krypton ($Z=36$) nuclei, with mean mass number of 105 to 108.

The fractions in Figure \ref{b:zs} were calculated for two different neutron stars, for $M_0=1.4~{\textnormal M_\odot}$ and $R_0=10$ km and a less compact neutron star with $M_0=1.0~{\textnormal M_\odot}$ and $R_0=20$ km. Although the crusts of these two different neutron stars are very different in size (see Figure \ref{b:mr}) the distributions of the nuclei are almost unaltered. This means that all existing nonaccreting cold neutron stars should exhibit a very similar composition of their outer crusts. Of course this conclusion is only valid, if the neutron star has actually reached the absolute ground state during its evolution and if one neglects contaminations of any fall-back supernova material.
\section{Summary}
For given neutron star configurations, the mass and thickness of the outer crust of nonaccreting cold neutron stars were calculated. For comparison, different theoretical nuclear mass tables were used as the basis for the equation of state. Only small deviations of the order of a few percent for the diameter of the outer crust were found for the different theoretical nuclear models, the masses were almost not affected at all. For a canonical neutron star with $M_0=1.4~{\textnormal M_\odot}$ the mass of the outer crust is typically of the order of $10^{-4}$ to $10^{-5}~{\textnormal M_\odot}$, with a radius of 300 to 800 m, which corresponds to a fraction of the total radius of 3 to 5 \%. It was shown, that the outer crust is the lighter and smaller, the more compact the neutron star is. Next, the composition of the outer crust was determined. For two of the models, NL3 and FRDM, a bimodal distribution in the mass number was found, reflecting the presence of the neutron magic numbers 50 and 82. Single nuclei of the sequence of groundstate nuclei can give a significant contribution to the overall composition, if they appear at high densities. Depending on the nuclear model, the mean mass number was found to be between 105 and 108, the mean charge number between 34 and 36. Finally it was noticed, that the composition is almost independent of the mass and the radius of the neutron star and thus is universal for all nonaccreting cold neutron stars under the assumptions made. 
\begin{acknowledgments}
Matthias Hempel gratefully acknowledges support from the Frankfurt Institute for Advanced Studies and the Helmholtz Research School for Quark Matter Studies. 
\end{acknowledgments}

\begin{widetext}

\begin{figure}[ht]
    \includegraphics[width=14cm]{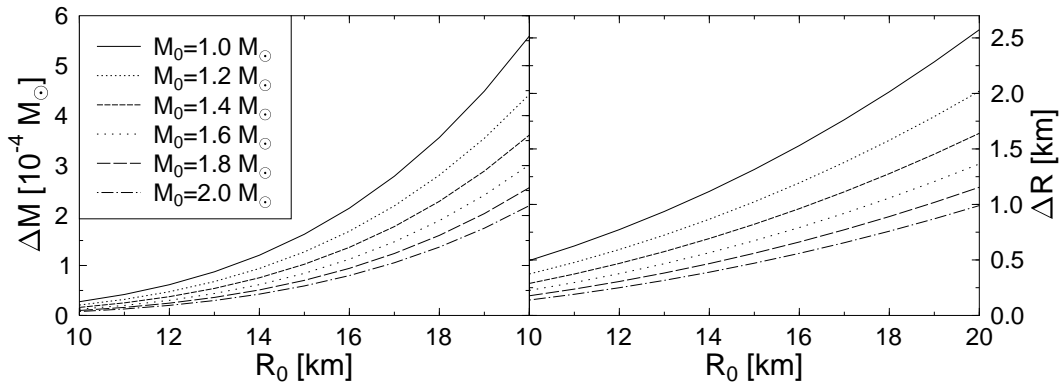}
    \caption{The mass $\Delta M$ and thickness $\Delta R$ of the outer crust in dependence of the inner radius $R_0$ for different inner masses $M_0$, calculated for the mass table FRDM. The initial pressure of the outer crust is $P_0=7.9\times 10^{29}$ dyne/cm$^2$.}
    \label{b:mr}
\end{figure}

\begin{figure}[ht]
    \includegraphics[width=14cm]{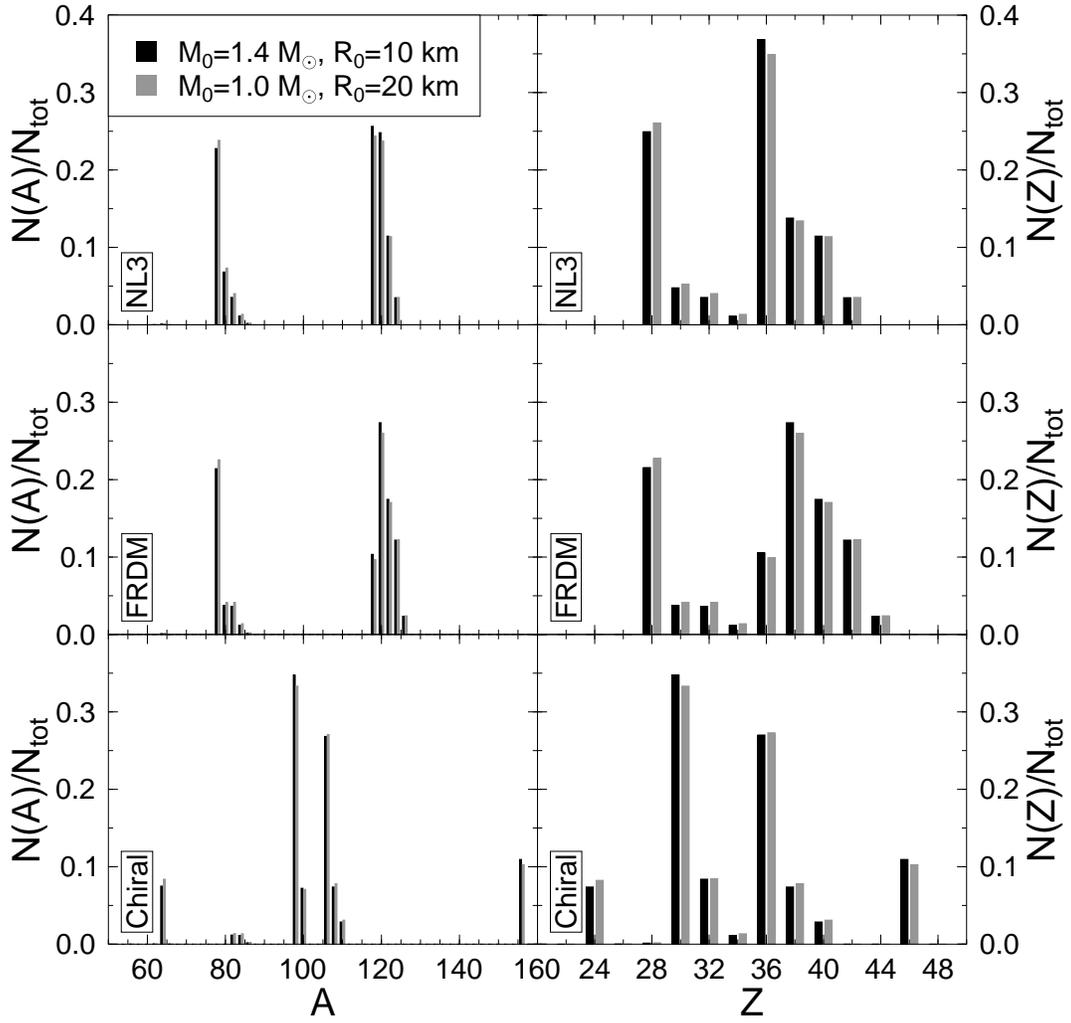}
    \caption{Distribution of the fraction of nuclei on the total number of nuclei in the outer crust $N_{tot}$ in dependence on the mass number $A$ (left side) and the charge number $Z$ (right side) for the equations of state NL3, FRDM, and Chiral. Black bars show the results for a neutron star with $M_0=1.4~{\textnormal M_\odot}$, $R_0=10$ km, gray ones for $M_0=1.0~{\textnormal M_\odot}$, $R_0=20$ km. The corresponding crusts are 288 m and 2,754~m thick with $1.61\times 10^{-6}~{\textnormal M_\odot}$ and $5.59\times 10^{-4}~{\textnormal M_\odot}$, respectively. $P_0=7.9\times 10^{29}$ dyne/cm$^2$.}
    \label{b:zs}
\end{figure}

\end{widetext}
\end{document}